\documentclass[twocolumn,showpacs,preprintnumbers,amsmath,amssymb]{revtex4}

\input epsf

\begin{document}

\title{The clustering of polarity reversals of the geomagnetic field}

\author{V. Carbone$^{1,2}$, L. Sorriso-Valvo$^{3}$, A. Vecchio$^4$, F. Lepreti$^{1,2}$, P. Veltri$^{1,2}$, P. Harabaglia$^5$, and I. Guerra$^1$}

\affiliation{
1) Dipartimento di Fisica, Universit\`a della Calabria, Ponte P. Bucci, Cubo 31C,
87036 Rende (CS), Italy \\
2) CNISM, Unit\`a di Cosenza, Rende (CS), Italy \\
3) Liquid Crystals Laboratory, INFM/CNR, Via P. Bucci 31/C I-87036 Rende(CS), Italy \\
4) Osservatorio Astrofisico di Arcetri, Largo E. Fermi 5, I-50125 Firenze, Italy \\
5) Dipartimento di Strutture, Geotecnica e Geologia Applicata all'Ingegneria, Universit\`a della Basilicata I-85100 Potenza, Italy.
}

\begin{abstract} 

Often in nature the temporal distribution of inhomogeneous stochastic point processes can be modeled as a realization of renewal Poisson processes with a variable rate. Here we investigate one of the classical examples, namely the temporal distribution of polarity reversals of the geomagnetic field. In spite of the commonly used underlying hypothesis, we show that this process strongly departs from a Poisson statistics, the origin of this failure stemming from the presence of temporal clustering. We find that a L\'evy statistics is able to reproduce paleomagnetic data, thus suggesting the presence of long-range correlations in the underlying dynamo process. 

\end{abstract}


\date{\today}

\maketitle

\begin{figure}
\epsfxsize=8cm 
\centerline{\epsffile{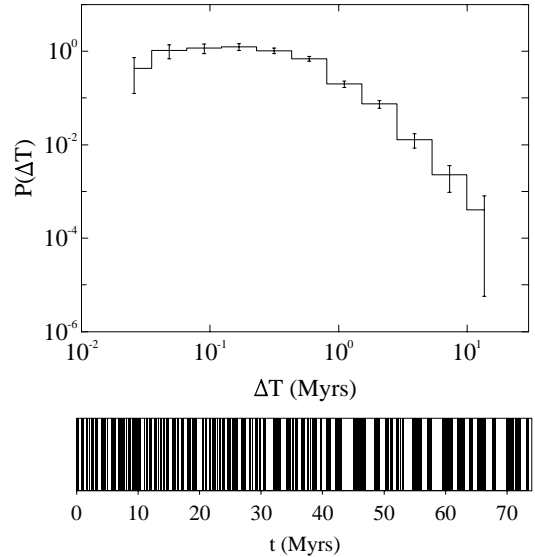}}
\caption{Upper panel: probability density function $P(\Delta t)$ of persistence times $\Delta t$ (statistical errors are shown as vertical bars). Lower panel: polarity of the earth's magnetic field (from today). The solid bar are the normal (present) polarity. We used the CK95 dataset.} 
\label{fig1}
\end{figure}

Local paleomagnetic measurements of the geomagnetic field \cite{generale,core,CK95} reveal a sequence of sudden and occasional global polarity reversals in the last $160$ million years. The magnetic dipole typically reverses its direction in a few $10^3$ years, while the time intervals between successive reversals range from $10^4$ up to $10^7$ years \cite{generale,CK95,valet93}. Even though polarity reversals can be ascribed to the Earth's magnetic dynamo \cite{generale,review,dynamo,stefani05}, details of the mechanism remain poorly understood. The debate on the origin of reversals, the modeling of the trigger (external or internal to Earth) and the generation of longer variations in the average reversal rate is still open (cfr. e.g. Ref.s \cite{core,yamazaki}). A look at the sequence of reversals, for example the most used CK95 database \cite{CK95}, reproduced in fig. \ref{fig1}, shows that polarity reversals seem to be the result of a non-periodic chaotic (or stochastic) process. Actually non regular reversals can be observed in the framework of purely deterministic toy models that mimic the dynamics of the dynamo effect with only few modes \cite{rikitake,crossley,turcotte}, in the framework of noise-induced switchings between two metastable states \cite{schmitt,hoyng02,hoyng04} or in mean-field dynamo with a noise-perturbed $\alpha$ profile \cite{giesecke,stefani05}. In principle geodynamo is described by 3D global Magnetohydrodynamics (MHD) that self-consistently solve for the fluid flow, thermodynamics and magnetic fields with all nonlinear feedbacks (for a review see Ref. \cite{review} and references therein, and the results of the recent $15.2$ TFlops simulation of geodynamo on the Earth Simulator \cite{earth}). Although some numerical codes have simulated short series of spontaneous reversals, none have been run so far at high enough resolution to be confident that the critical dynamics is being captured in the simulation. 
%%%%%%%%%%%%%%%%%%%%%%%%%%%%%%%%%%%%%%%%%%%%%%%%%%%%%%%%%%%%%%%%%%%%%%%%%
%%%%%%%%%%%%%%%%%%%%%%%%%%%%%%%%%%%%%%%%%%%%%%%%%%%%%%%%%%%%%%%%%%%%%%%%%

In spite of the paucity of data sets, it is commonly assumed that the phenomenon of reversals stems from an underlying Poisson process. This conjecture is based on the fact that the polarity persistence times $\Delta t$, defined as the time intervals between two consecutive reversals $\Delta t = t_{i+1}-t_i$, seem to be exponentially distributed \cite{generale,mcfadden,hoyng02,constable}, namely $P(\Delta t) \sim \lambda \exp(-\lambda \Delta t)$ where $\lambda$ represents the rate of occurrence of reversals. Different analyses (cfr. e.g Ref. \cite{jonkers}) state that the frequency distribution of intervals between Cenozoic geomagnetic reversals approximate a power law for large $\Delta t$. Even a rough look at the probability distribution function $P(\Delta t)$ (cfr. fig \ref{fig1}) shows that the situation is not clear, mainly in presence of a poor data set  with high statistical errors.

Recently Constable \cite{constable} raised two main statistical features to the attention of the  scientific community, namely: i) the distribution of events shows a paucity of short intervals; ii) the rate of occurrence of events is time-dependent $\lambda = \lambda(t)$. 
The author investigated the temporal distribution of $\lambda(t)$, showing that a Poisson model with a monotonic rate, either increasing or decreasing, is not a good model for the reversal process. On the contrary, reversals could be perhaps modeled as a renewal Poisson process with a rate that must change sign at some interval before $158$ My \cite{constable}. 
In any case, modeling $\lambda(t)$ over the entire time interval is not an easy task (cfr. also Ref.s \cite{altripoisson,mcfadden}). Moreover, in this situation, namely when the rate $\lambda$ depends on time, the distribution of persistence times remains without physical meaning and cannot be used to determine the Poisson character of events \cite{feller}. 

The problem is perhaps more general than what we present here, because abrupt flow reversals have been found also in the large-scale circulation during turbulent Rayleigh-Benard convection \cite{benzi}, or in the wind direction in atmosphere \cite{atmosfera}. The conjecture that reversals are Poisson events is made in all cases. Here, starting from the above experimental evidences, and using a simple statistical test on some databases, we will investigate whether a conjecture based on the occurrence of a Poisson process for reversals is correct or not. We will show that this is not the case, and that geomagnetic reversals are clustered in time, a result which stems from the presence of memory in the process generating polarity reversals. 
%%%%%%%%%%%%%%%%%%%%%%%%%%%%%%%%%%%%%%%%%%%%%%%%%%%%%%%%%%%%%%%%%%%%%%%%%
\begin{figure}
\epsfxsize=10cm 
\centerline{\epsffile{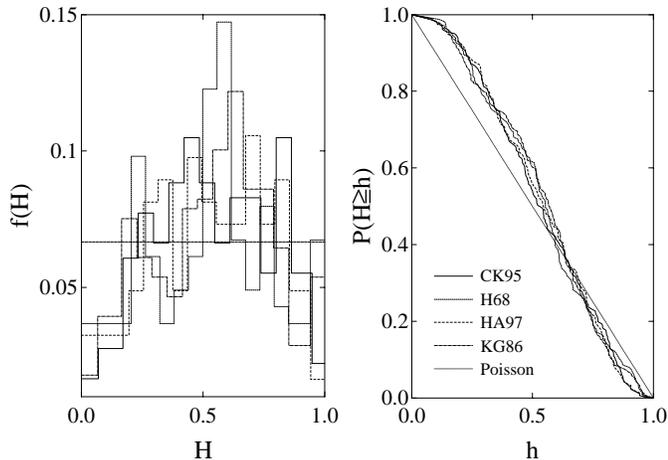}}
\caption{The probability densities $f(H)$ of the stochastic variable $H$ (left panel) and the cumulative probability $P(H \geq h$ (right panel), are presented for all datasets. Theoretical probabilities observed under a Poisson statistics are also shown.} 
\label{fig2}
\end{figure}
%%%%%%%%%%%%%%%%%%%%%%%%%%%%%%%%%%%%%%%%%%%%%%%%%%%%%%%%%%%%%%%%%%%%%%%%%

Under the experimental evidence that the rate of reversals is not constant, we can test, as a zero-th order hypothesis, wheter an approach based on the occurrence of a \textit{Local Poisson Process} is correct or not. In other word, according to Constable \cite{constable} it can be conjectured that (hypothesis $H_0$), even if it cannot be decide wheter globally the reversals stem from a Poisson process, an underlying time-varying Poisson process could be assumed to originate the geomagnetic reversals. Since the reversals rate $\lambda(t)$ is not known, we have to build up a test which is \textit{independent} on the rate $\lambda$. This can be done through a measure used previously in the framework of solar flares \cite{boffetta,lepreti}. We introduce a statistical quantity that is nothing but the suitably normalized local time interval between reversals. Let us consider the time sequence of reversals as a point-like process and suppose that each reversal occurs at the discrete time $t_i$. Let $\delta t_i = \min \{t_{i+1}-t_i;t_i-t_{i-1}\}$ and let $\tau_i$ be either $\tau_i = t_{i-1}-t_{i-2}$ (if  $\delta t_i = t_i-t_{i-1}$) or $\tau_i = t_{i+2}-t_{i+1}$ (if  $\delta t_i = t_{i+1}-t_i$), so that $\delta t_i$ and $\tau_i$ are the two persistence times following or preceeding a given reversal at $t_i$. If the local Poisson hypothesis $H_0$ holds, both $\delta t_i$ and $\tau_i$ are independently distributed according to an exponential probability density: $p(\delta t) = 2 \lambda_i \exp(-2 \lambda_i \delta t)$ and $p(\tau)= \lambda_i \exp(-\lambda_i \tau)$ with local rate $\lambda_i$. Then, under the hypothesis $H_0$, the stochastic variable $H$, defined as 

\begin{equation}
H(\delta t_i, \tau_i) = \frac{2 \delta t_i}{2 \delta t_i + \tau_i}
\label{acca}
\end{equation}
is uniformly distributed in $[0;1]$ and has cumulative distribution $P(H \geq h)  = 1-h$, independent on $\lambda_i$ \cite{lepreti}. In a process where $\tau_i$s are systematically smaller than $2 \delta t_i$s, clusters are present and the average value of $H$ is greater than $1/2$. On the contrary when the underlying process is characterized by voids, the average value of $H$ is less than $1/2$. From time series, it is easy to calculate the probability $P(H \geq h)$ and the probability density function $f(H)$. 

We apply the above test to four different sequences of geomagnetic polarity reversals, namely to the already mentioned CK95, to H68, HA97 and KG86 \cite{datasets}. We calculate the probability density function $f(H)$ reported in figure \ref{fig2} (left panel). As can be seen, a significant deviation from the uniform distribution is evident in all datasets, the departure of polarity reversals from local Poisson statistics stemming from the presence of clusters. Then a clear deviation of the observed cumulative probability $P(H \geq h)$ from a linear law (cfr. fig. \ref{fig2} right panel), expected under $H_0$, is obtained. A Kolmogorov-Smirnov (KS) test applied to the cumulative distributions confirms that the assumed hypothesis $H_0$ is not reliable and must be rejected (the significance level of the KS test being smaller than $0.5 \%$ for all datasets). 
%%%%%%%%%%%%%%%%%%%%%%%%%%%%%%%%%%%%%%%%%%%%%%%%%%%%%%%%%%%%%%%%%%%%%%%%%
\begin{figure}
\epsfxsize=7cm 
\centerline{\epsffile{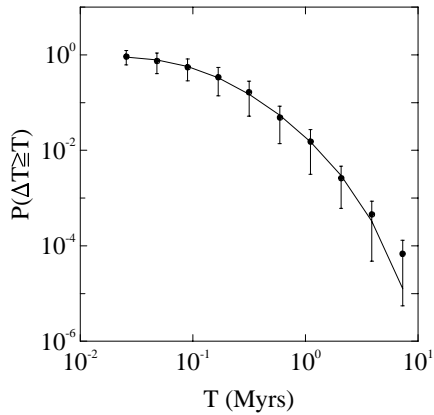}}
\caption{The cumulative probability $P(\Delta t \geq T)$ as a function of $T$ (symbols) obtained from the CK95 database (vertical bars represent statistical errors). The full line represents the fit obtained with the truncated L\'evy function (\ref{koponen1}). The best-fit parameters are $\nu=1.09\pm 0.1$ and $\theta = 0.15 \pm 0.04$.} 
\label{fig_levy}
\end{figure}
%%%%%%%%%%%%%%%%%%%%%%%%%%%%%%%%%%%%%%%%%%%%%%%%%%%%%%%%%%%%%%%%%%%%%%%%%

To further characterize the origin of the departure of polarity reversals from a local Poisson process, we try to describe the statistics of persistence times with a L\'evy process. The L\'evy functions are stable, and can be obtained from the Central Limit Theorem by relaxing the hypothesis of finite variance \cite{levy}. To avoid problems arising from the infinite variance, which is unlikely in real physical processes, a Truncated L\'evy Flight distribution (TLF) has been proposed~\cite{mantegna} introducing a cutoff in the standard L\'evy. The process can be generated by a random variable $x$ according to the distribution function  $f(x) \sim e^{-\theta\mid x\mid}\mid x\mid^{-1-\nu}$, where $\theta$ represents the cutoff rate and $\nu$ is the characteristic exponent. It can be shown~\cite{koponen} that the limit distribution $P(z)$ of the sum $z$ of random variables $x$ can be computed, for $\nu\neq 1$, by the following integral 

\begin{eqnarray}
P(z) &=& C{\int_0}^\infty dk\cos(zk) \exp\left\{\frac{\theta^\nu}{\cos(\pi \nu/2)}- 
\right. \nonumber \\
&-& \left.
\frac{2\pi(z^2+\theta^2)^{\nu/2}\cos[\nu\arctan(\mid
z\mid/\theta)]}{\nu\Gamma(\nu)\sin(\pi\nu)}\right\}
\label{koponen1}
\end{eqnarray}
where $C$ is a normalization factor and $\Gamma(\nu)$ is the usual gamma function. For $\nu\ge 2$ we recover a normal random process, the result of the integration being a Gaussian PDF, while smaller values of $\nu$ represent increasing deviation from Gaussianity. 
We thus fitted the measured cumulative PDF of persistences $P(\Delta t \geq T)$, computed as described above and reported in figure~\ref{fig_levy}, with equation (\ref{koponen1}). The equation has been numerically integrated using a standard minimum-$\chi^2$ procedure. We obtained the best-fit parameters $\nu=1.09\pm 0.1$, and $\theta = 0.15 \pm 0.04$ with a reduced $\chi^2 \simeq 0.5$. 

The possibility of reproducing with a L\'evy function the cumulative distribution of persistence times indicates both that the process underlying the polarity reversals is statistically self-similar in time, and that a certain amount of memory, due to long-range correlations, is present in the process. 
From a physical point of view we could expect that the dynamics of the fluid earth core is affected by its hystory, thus generating correlations among reversals. This feature has recently be observed for the solar dynamo \cite{solardyn}. Of course a lot of different random processes can reproduce the departure from a Poisson statistics. Only the next generation of simulators will be able to produce datasets with enough reversals to allow us toinvestigate in detail the occurrence of long-range correlations in 3D global MHD simulations of geodynamo. In this framework, statistical analysis on real data, among other, play the key role of discriminating among different random processes that can reproduce the departure from poisson statistics, thus increasing our knowledge of the geodynamo process. Dynamical models can help us because they describe, with only few physical ingredients, some gross features of the enormous complexity of the geodynamo process. In this perspective it is useful to compare the statistics of reversals observed in toy models with statistics obtained on real datasets.
%%%%%%%%%%%%%%%%%%%%%%%%%%%%%%%%%%%%%%%%%%%%%%%%%%%%%%%%%%%%%%%%%%%%%%%%%
\begin{figure}
\epsfxsize=10cm 
\centerline{\epsffile{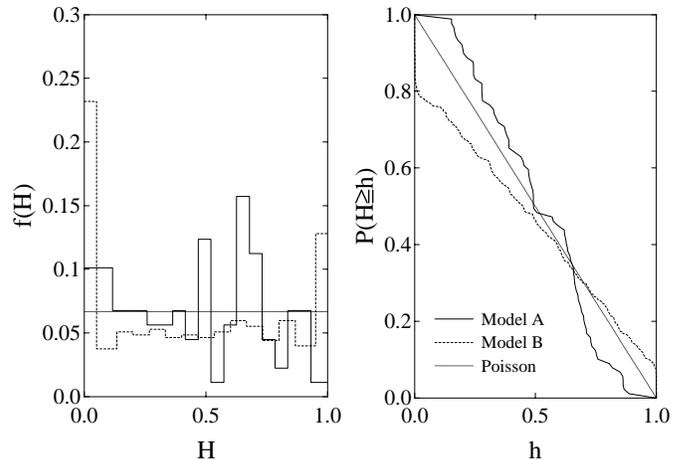}}
\caption{The probability densities $f(H)$ of the stochastic variable $H$ (left panel) and the cumulative probability $P(H \geq h$ (right panel), are presented for both models A and B. Theoretical probabilities observed under a Poisson statistics are also shown.} 
\label{fig_models}
\end{figure}
%%%%%%%%%%%%%%%%%%%%%%%%%%%%%%%%%%%%%%%%%%%%%%%%%%%%%%%%%%%%%%%%%%%%%%%%%

As an example, we investigate the sequence of random reversals generated by a standard numerical analysis of the two-disk chaotic geodynamo model \cite{rikitake}. The model, known as ``Rikitake dynamo", is described by the following ordinary differential equations: $dx/dt = -x + yz$, $dy/dt = -y + x(z-15/4)$, and $dz/dt = 1-xy$ (in the following model A). The cumulative distribution of $H$, obtained by the times of reversals of the variable $x(t)$, has been reported in figure \ref{fig_models}. It is evident that the chaotic dynamics of the model is responsible for the presence of correlations among the reversals. As a further example we investigate a dynamical model \cite{hoyng04} based on stochastic excitation of the axisymmetric component of the magnetic field (model B), that reads: $dx/dt = (1-x^2)x + V_{11} x + V_{12} y + V_{13} z$, $dy/dt = -ay-cz + V_{21} x + V_{22} y + V_{23} z$, and $dz/dt = cy-az + V_{31} x + V_{32} y + V_{33} z$. The $V_{ij}(t)$ are $9$ independent random functions of time, with zero mean and equal r.m.s. magnitude, that are renewed after a time $\tau_c$. We use the standard parameters $a = 2$, $c = 5$, $\tau_c = 0.01$, and $D = \left< V_{ij}^2\right> \tau_c$, where $D = 0.4$ \cite{hoyng04}. The cumulative distribution of $H$ (fig. \ref{fig_models}), calculated by using the times of the random reversals of the variable $x(t)$, has been reported in figure \ref{fig_models}. Even in this case a departure from a Poisson distribution is observed, due to the correlations introduced through $\tau_c$. 

Even if the chaotic dynamics within the Rikitake dynamo roughly reproduces the presence of clustering among reversals, a look at fig.s \ref{fig_models} reveals that in model A the departure from the local Poisson distribution seems to be affected by the presence of sudden jumps in the cumulative distribution of $H$. Say the reversal time series $x(t)$ presents few recurrent persistence times of equal extent.
On the other hand the model B shows a different departure from Poisson statistics, that is mainly due to both very low values and, even if in minor extent, to very high values of $H$. The time behaviour of reversals in the model B is then hardly compatible with the clustering process evidenced through our analysis. Further toy models based on completely random external triggers of reversals (e.g. stochastic resonance) cannot describe the geodynamo process. 

As a conclusion, we investigated the statistics of persistence times between geomagnetic reversals. We applied a statistical test, showing that geomagnetic reversals stem from an underlying process that is far from being locally Poissonian, as currently conjectured \cite{constable}. A L\'evy function is able to nicely fit the probability distribution of persistence times. Although not investigated up to now in a geophysical framework, our results can be interpreted as a strong evidence for the presence of correlations between reversal events. These correlations arise from some degree of memory in the underlying geodynamo process \cite{valet,stefani05} that gives rise to clustering of reversals. 

\acknowledgments{We acknowledge two anonymous Referees whose comments have improved the final version of the paper.}

\end{document}